\documentstyle[11pt]{article}

\makeatletter
\@addtoreset{equation}{section}
\makeatother

\def\dalemb#1#2{{\vbox{\hrule height .#2pt
        \hbox{\vrule width.#2pt height#1pt \kern#1pt
                \vrule width.#2pt}
        \hrule height.#2pt}}}
\def\square{\mathord{\dalemb{6.8}{7}\hbox{\hskip1pt}}}

\def\td{\tilde}

\let\a=\alpha    
    
\let\l=\lambda \let\m=\mu

\def\nn{\nonumber} \def\bd{\begin{document}} \def\ed{\end{document}}
\def\ds{\documentstyle} \let\fr=\frac \let\bl=\bigl \let\br=\bigr
\let\Br=\Bigr \let\Bl=\Bigl
\let\bm=\bibitem
\let\na=\nabla
\let\pa=\partial \let\ov=\overline
\newcommand{\be}{\begin{equation}}
\newcommand{\ee}{\end{equation}}
\def\ba{\begin{array}}
\def\ea{\end{array}}
\def\ft#1#2{{\textstyle{{\scriptstyle #1}\over {\scriptstyle #2}}}}
\def\fft#1#2{{#1 \over #2}}
\def\del{\partial}
\def\ie{{\it i.e.\ }}
\def\sst#1{{\scriptscriptstyle #1}}
\def\oneone{\rlap 1\mkern4mu{\rm l}}
\def\Z{\rlap{\sf Z}\mkern3mu{\sf Z}}
\def\R{\rlap{\rm I}\mkern3mu{\rm R}}
\def\ss{{Scherk-Schwarz\ }}
\def\cramp{\medmuskip = 2mu plus 1mu minus 2mu}
\def\crampest{\medmuskip = 1mu plus 1mu minus 1mu}
\def\uncramp{\medmuskip = 4mu plus 2mu minus 4mu}
\newcommand{\ho}[1]{$\, ^{#1}$}
\newcommand{\hoch}[1]{$\, ^{#1}$}
\newcommand{\bea}{\begin{eqnarray}}
\newcommand{\eea}{\end{eqnarray}}
\newcommand{\ra}{\rightarrow}
\newcommand{\lra}{\longrightarrow}
\newcommand{\Lra}{\Leftrightarrow}
\newcommand{\ap}{\alpha^\prime}
\newcommand{\bp}{\tilde \beta^\prime}
\newcommand{\tr}{{\rm tr} }
\newcommand{\Tr}{{\rm Tr} }
\newcommand{\NP}{Nucl. Phys. }
\newcommand{\tamphys}{\it Center for Theoretical Physics,
Texas A\&M University, College Station, Texas 77843}
\newcommand{\ens}{\it Laboratoire de Physique Th\'eorique de l'\'Ecole
Normale Sup\'erieure\hoch{2}\\
24 Rue Lhomond - 75231 Paris CEDEX 05}

\newcommand{\auth}{I.V. Lavrinenko{\hoch{\dagger}}, 
H. L\"u{\hoch{\ddagger}} and 
C.N. Pope{\hoch{\dagger1}}}

\begin{document}

\hfill{CTP-TAMU-43/97}

\hfill{LPTENS-97/50}

\hfill{hep-th/9710243}

\hfill{Oct. 1997}

\vspace{20pt}

\begin{center}
{ \large {\bf Fibre Bundles and Generalised Dimensional Reductions}}

\vspace{30pt}
\auth

\vspace{15pt}
{\hoch{\dagger}\tamphys}

\vspace{10pt}
{\hoch{\ddagger}\ens}

\vspace{40pt}

\underline{ABSTRACT}
\end{center}

     We study some geometrical and topological aspects of the
generalised dimensional reduction of supergravities in $D=11$ and
$D=10$ dimensions, which give rise to massive theories in lower
dimensions.  In these reductions, a global symmetry is used in order
to allow some of the fields to have a non-trivial dependence on the
compactifying coordinates.  Global consistency in the internal space
imposes topological restrictions on the parameters of the
compactification as well as the structure of the space itself.
Examples that we consider include the generalised reduction of the
type IIA and type IIB theories on a circle, and also the massive
ten-dimensional theory obtained by the generalised reduction of $D=11$
supergravity.

\vspace{40pt}

{\vfill\leftline{}\vfill
\vskip  10pt
\footnoterule
{\footnotesize  \hoch{1} Research supported in part by DOE 
Grant DE-FG03-95ER40917 \vskip  -12pt} \vskip 14pt
{\footnotesize
        \hoch{3} Unit\'e Propre du Centre National de la Recherche
Scientifique, associ\'ee \`a l'\'Ecole Normale Sup\'erieure \vskip -12pt}
                       \vskip 10pt
{\footnotesize \hoch{\phantom{3}} et \`a l'Universit\'e de Paris-Sud 
\vskip -12pt}}

\pagebreak
\setcounter{page}{1}

\section{Introduction}

    The standard Kaluza-Klein procedure for dimensional reduction on a
circle or a torus consists of two elements.  Firstly, one performs a
Fourier expansion of all the higher-dimensional fields, in terms of
the mode functions on the circle or torus.  Then, one usually follows
this by a truncation in which all but the zero mode sectors are
discarded.  The two steps can equivalently be performed
simultaneously, by taking the lower-dimensional fields to be
independent of the compactification coordinates.  This procedure is
guaranteed to be consistent, \ie all solutions of the
lower-dimensional theory will also be solutions of the original
higher-dimensional one.  This can be seen by performing the reduction
and truncation at the level of the equations of motion of the
higher-dimensional theory.  In general, the consistency of a
truncation is threatened if non-linear terms involving the fields that
are retained can generate mode-function dependences of fields that are
being truncated to zero.  Clearly this cannot happen here, since one
cannot generate non-zero Fourier modes from products of zero modes.
Thus the truncation is consistent.

    What is less familiar is that in circle or torus compactifications
there can be slightly more general truncations that are also
consistent, even when certain dependences on the compactification
coordinates are allowed.  Suppose, for example, we consider a
reduction on a single circle with coordinate $z$.  If a particular
field $\Phi$ appears in the higher-dimensional equations of motion
only {\it via} its derivative, then it follows that the usual
Kaluza-Klein truncation $\hat \Phi(x,z)= \Phi(x)$ can be generalised
to $\hat\Phi(x,z) = \Phi(x) + m\, z$, where $m$ is an arbitrary
constant.  There will still be no $z$ dependence in the
higher-dimensional equations of motion, and consequently the
truncation to the lower-dimensional theory will still be consistent.
A typical simple situation where such a generalised reduction can be
performed is when the field $\Phi$ is an axion.

     In fact the above example is still not the most general kind of
reduction that one may consider.  The essential feature in that
example was that the field $\Phi$ was subject to the shift
symmetry $\Phi\rightarrow \Phi + c$, where c is a constant.  This is a
global $\R$ symmetry.  One could still, for example, perform a
generalised reduction even in a different choice of field variables,
related to the original ones by local transformations, in which the
field $\Phi$ was no longer covered everywhere by a derivative.  All
that is necessary is that the theory admit the global $\R$ symmetry.
The generalised reduction could then be effected by performing a 
global symmetry transformation with a parameter $c$ that is replaced by the
quantity $m\, z$, and then following this with a ``standard''
($z$-independent) Kaluza-Klein reduction.  The symmetry under constant
shifts $c$ would then ensure that after setting $c=m\, z$, only terms
where a derivative falls on $z$ would survive, and thus the
transformed reduction scheme would still give $z$ independence,
hence ensuring the consistency of the truncation.  In general, the
resulting lower-dimensional theory will be a {\it massive} theory,
with masses proportional to the parameter $m$ 
\cite{ss1,bdgpt,clpst,lpdomain,llp,classp}.

     Further generalisations of the above scheme are clearly
possible, in which one considers {\it any} global symmetry of the
higher-dimensional equations of motion, and then replaces the constant
parameters of the transformations by terms linearly dependent on the
compactification coordinates on the torus or circle.  For example, we
could consider the $SL(2,\R)$ global symmetry of the type IIB theory,
and perform a generalised reduction on a circle, by first performing an
$SL(2,\R)$ transformation in which the three parameters are allowed to be
linearly dependent on the compactification coordinate.  Of course
there will really only be two, rather than three, independent mass
parameters in the nine-dimensional theory, since transformations in
the stability subgroup of $SL(2,\R)$ will not have any effect.
Another example of this kind is a generalised circle reduction of the type
IIA theory.  This theory has a global $\R$ symmetry of the action,
corresponding, in the usual field variables, to a shift of the dilaton
accompanied by appropriate rescalings of the other fields.  (One can,
in this case, redefine fields to absorb all the non-derivative dilaton
dependence, but, as we have remarked above, it is not essential to do
this in order to be able to perform the generalised reduction.)

    There are also global scaling symmetries in supergravities that
lie outside the Cremmer-Julia symmetries \cite{cj3,cj2} that are
usually considered (see also \cite{ms}).
Specifically, there is a constant scaling symmetry that leaves the
equations of motion invariant, although it is not a symmetry of the
action since it scales it uniformly by an overall constant
factor.\footnote{The significance of this scaling symmetry for
understanding the lower-dimensional Cremmer-Julia symmetries was
recently observed in \cite{cjlp}.   It was also used in
\cite{trombone} to construct spectrum-generating symmetries for the
BPS solitons, and in the rheonomy approach to
supersymmetry; see, for example \cite{cdf}.}  (In fact this feature,
of leaving only the equations of motion invariant, arises also for the
Cremmer-Julia symmetries in even dimensions.)  The scaling symmetry is
the generalisation of the scaling symmetry of the pure Einstein theory
under $g_{\mu\nu} \rightarrow \lambda^2\, g_{\mu\nu}$, where $\lambda$
is a constant.  For example, in eleven-dimensional supergravity the
equations of motion are invariant under constant field scalings which,
in the bosonic sector, take the form
\be
g_{\mu\nu} \longrightarrow \lambda^2\, g_{\mu\nu}\ ,\qquad
A_{\mu\nu\rho} \longrightarrow \lambda^3\, A_{\mu\nu\rho}\ .
\ee
This global $\R$ symmetry too can be used to allow generalised
Kaluza-Klein reductions.
 
     In this paper, we shall explore a number of aspects of the
generalised reduction of supergravity theories.  We begin in section 2
by considering some simple examples involving generalised reductions
for axionic scalars.  Our discussion also addresses some geometrical
and topological aspects of the internal compactifying spaces.
Although we shall be restricting attention to compactifications on
circles and products of circles, it can nevertheless be that
topologically non-trivial ``twists'' can arise, due to the imposition
of the generalised reduction ansatz.  Thus, as we shall show, the
total internal compactifying space can sometimes have the structure of
a non-trivial torus bundle over a base torus.
In particular, we show how, in the cases where the internal space is a
torus bundle, there are topologically-imposed relations between the
compactification periods of the different circles.  These in turn
imply relations between the various mass parameters.  

    In section 3 we present a general discussion of more general
reduction procedures, in which fields other than axions are allowed
non-trivial dependence on the compactifying coordinates.  In section
4, we give some examples of this kind, for theories with global
symmetries that include shifting symmetries of the dilaton.  Our
examples here include the generalised reduction of type IIA to $D=9$,
using the dilatonic $\R$ symmetry, and generalised reductions using
the global $SL(2,\R)$ symmetries of the type IIB theory in $D=10$ and
the type IIA theory in $D=9$.

    In section 4, we consider the generalised reduction of
eleven-dimensional supergravity to $D=10$, using the global scaling
$\R$ symmetry.  This gives rise to a massive type IIA supergravity in
$D=10$.  In fact this theory was also considered recently in \cite{hlw},
where it was obtained by different, although not entirely unrelated,
techniques.  It is, unfortunately, not the same as the massive IIA
theory found by Romans \cite{r}, which has found a r\^ole in recent
times in connection with D8-branes in string theory \cite{pw}.  Nonetheless,
as we shall show, the new theory has its own interesting features,
including the fact that it admits a ten-dimensional de Sitter
spacetime solution.

The paper ends
with concluding remarks in section 6.  Some details of Kaluza-Klein
reductions are included in an appendix.

\section{Domain walls and torus bundles}

\subsection{Generalised reductions with axionic scalars}

    In this section, we shall consider domain walls that arise as solutions 
of the massive theories obtained by the generalised Kaluza-Klein 
reduction of eleven-dimensional supergravity.  We shall concentrate on cases 
where the massive theories are obtained by generalised toroidal reductions, 
The simplest example is provided by the \ss reduction of maximal 
supergravity in nine dimensions, giving a massive theory in $D=8$.  The 
nine-dimensional theory has one axion, namely ${\cal A}_0^{(12)}$, which can 
be reduced to eight dimensions using the generalised Kaluza-Klein ansatz
\be
{\cal A}_0^{(12)}(x,z_3) = m\, z_3 + {\cal A}_0^{(12)}(x)\ . \label{d9red}
\ee
One can easily see from the structure of the Chern-Simons modifications to 
the field strengths in $D=9$, given in (\ref{csterm}), that 
after suitable field 
redefinitions this axion can be covered by a derivative everywhere, and thus 
the reduction to $D=8$ will be consistent since no $z_3$ dependence will 
occur in the nine-dimensional equations of motion.  This means that one can 
consistently extract eight-dimensional equations of motion, whose solutions 
will all be solutions of the nine-dimensional theory, and hence also of the 
eleven-dimensional theory.  The resulting eight-dimensional theory is
maximally supersymmetric, with mass terms for certain of the gauge 
potentials, and in addition a cosmological term of the form \cite{bdgpt}
\be
{\cal L}_{\rm cosmo} = -\ft12 e\, m^2\, e^{\vec b_{123}\cdot\vec\phi}\ ,
\ee
where $\vec b_{123}$ is a constant 3-vector characterising the couplings of 
the three dilatonic scalars $\vec \phi=(\phi_1,\phi_2,\phi_3)$ coming from 
the diagonal components of the compactifying metric \cite{lpsol,bdgpt}. 

    One solution of the massive theory that is of particular interest
is a domain wall, which is 
effectively like the ground state of the massive eight-dimensional theory.
(The massive theory admits no maximally-symmetric ground state, \ie neither 
Minkowski spacetime nor an anti de Sitter spacetime.)  The domain-wall
solution is given by
\bea
ds_8^2 &=& H^{1/6}\, dx^\mu\, dx_\mu + H^{7/6}\, dy^2\ ,\label{dw8met}\\
e^\phi &=& H^{-a/2}\ ,\qquad H= 1+ m|y|\ ,\nn
\eea
where $a\phi=\vec b_{123}\cdot\vec\phi$ and $a=\sqrt{19/3}$.
This is a BPS-saturated solution, which preserves $\ft12$ of the 
supersymmetry.  It can be viewed as the generalisation of the usual 
BPS-saturated $p$-brane solitons to the case where the field strength 
supporting the solution is a 0-form, namely the constant $m$, carrying a 
magnetic charge.

     When oxidised back to $D=11$, the metric becomes
\be
ds_{11}^2 = dx^\mu\, dx_\mu + H\, dy^2 + H\, (dz_2^2 + 
dz_3^2) + H^{-1}\, (dz_1 +m\, z_3\, dz_2)^2\ .\label{dw11}
\ee
The three-dimensional manifold on which the eleven-dimensional theory is 
compactified is described by the coordinates $(z_1,z_2,z_3)$.  It has the 
structure of a $U(1)$ principal fibre bundle, with coordinate $z_1$, over a 
two-dimensional base torus with coordinates $(z_2,z_3)$.  The bundle is 
non-trivial, since there is a
connection form ${\cal A}_1^{(1)}= m\, z_3\, dz_2$ on the fibre, which can be 
seen to be non-trivial since its curvature ${\cal F}_2^{(1)}=d{\cal 
A}_1^{(1)} = m\, dz_2\wedge dz_3$ has a non-vanishing integral over the 
volume of the 2-torus.   This imposes a periodicity condition on $z_1$, which 
we shall discuss in more detail later.

     At a fixed value of the transverse-space coordinate $y$, the internal 
metric in (\ref{dw11}) is of the form
\be
ds^2 = \lambda^2\, (dz_2^2 + dz_3^2) +\lambda^{-2}\, (dz_1 + m\, z_3 \, 
dz_2)^2 \ ,\label{bianchi2}
\ee
where $\lambda$ is a constant.  This metric is homogeneous, as can easily be 
seen from the fact that the vielbeins $e^1=\lambda^{-1}\, (dz_1+ m\, z_3\, 
dz_2)$, $e^2=\lambda\, dz_2$, $e^3=\lambda\, dz_3$ satisfy
\be
de^1 =-m\, \lambda^{-3}\, e^2\wedge e^3\ ,\qquad de^2=0\ ,\qquad
de^3=0 \ ,\label{b2struct}
\ee
and thus the structural coefficients $C_{bc}{}^a$ in the exterior 
derivatives $de^a=-\ft12 C_{bc}{}^a\, e^b\wedge e^c$ are constants.  The 
spin connection and curvature 2-form are given by
\bea
\omega_{12} &=& -\ft12 m\lambda^{-3}\, e^3\ ,\qquad 
\omega_{23}= \ft12 m\lambda^{-3}\, e^1\ ,\qquad 
\omega_{31}= -\ft12 m\lambda^{-3}\, e^2\ , \\
\Theta_{12} &=& \ft14 m^2\, \lambda^{-6}\, e^1\wedge e^2\ ,\qquad
\Theta_{23} = -\ft34 m^2\, \lambda^{-6}\, e^2\wedge e^3\ ,\qquad
\Theta_{31} = \ft14 m^2\, \lambda^{-6}\, e^3\wedge e^1 .\nn
\eea
In fact, the metric (\ref{bianchi2}) is precisely of the form of the Bianchi 
II metrics in the standard classification of three-dimensional homogeneous 
spaces \cite{classf}.   
All solutions of the eight-dimensional massive theory will 
preserve the topological structure of this three-dimensional internal space.

     Many further examples of massive supergravities coming from
M-theory can be found in all dimensions $D\le 8$
\cite{clpst,lpdomain,llp,classp}.  First of all, one can perform a
Scherk-Schwarz reduction on any one of the axions that arises in any
of the usual maximal supergravities in $D\le9$. A natural
generalisation is then to consider cases where $N\ge2$ axions are
simultaneously subjected to the Scherk-Schwarz reduction procedure.
In order to be able to do this, it is necessary that they can be
simultaneously covered everywhere by derivatives. In other words,
there should be at least an $R^N$ symmetry describing the simultaneous
global shift symmetries of the $N$ axions.  However, it is not
guaranteed that every such massive theory will admit a domain wall
solution.  The situation is analogous to that for multi-charge
$p$-brane solutions using field strengths of degree $\ge 1$, where
only certain combinations of field strengths may be used in their
construction \cite{lpsol,lpmulti}.  In both the present case where
the field strengths have degree $n=0$, and the more usual cases with
field strengths of degree $n\ge1$, the criterion for the existence of
multi-charge solutions is the same, namely that they exist if the set
of $N$ participating fields strengths have dilaton vectors $\vec c_\a$
($\a=1,\ldots, N$) that satisfy \cite{lpsol} 
\be 
\vec c_\a\cdot\vec c_\beta
=4\delta_{\a\beta} -\fft{2(n-1)(D-n-1)}{D-2}\ .  
\ee 
In the present context, every set of cosmological terms whose dilaton
vectors satisfy \cite{classp}
\be 
\vec c_\a\cdot\vec c_\beta =4\delta_{\a\beta}+\fft{2(D-1)}{D-2}
\label{dot} 
\ee 
can be arrived at by an appropriate Scherk-Schwarz
reduction on multiple axions.  Other cases where simultaneous
Scherk-Schwarz reductions are possible, but the resulting dilaton
vectors for the cosmological terms do not satisfy (\ref{dot}),
correspond to massive theories that do not admit any domain wall
solutions (nor do they admit Minkowski or anti de-Sitter solutions).

     An example of a massive theory involving simultaneous
Scherk-Schwarz reductions was presented in \cite{bdgpt}, where it was
shown that three of the four axions in the massless eight-dimensional
theory, for example ${\cal A}_0^{(12)}$, ${\cal A}_0^{(13)}$ and
$A_0^{(123)}$, can be simultaneously covered by derivatives.  Using
this $R^3$ symmetry, a massive supergravity in $D=7$ was constructed,
whose ``cosmological terms'' are of the form
\be
{\cal L}_{\rm cosmo}=-\ft12 m_1^2\, e^{\vec b_{124}\cdot\vec\phi} -
\ft12 m_3^2\, e^{\vec a_{1234}\cdot\vec\phi} -\ft12 (m_2 - m_1\, {\cal 
A}_0^{(23)} )^2\, e^{\vec b_{134}\cdot\vec\phi}\ .
\ee
If $m_1$ is non-zero, the redefinition ${\cal A}_0^{(23)} \rightarrow {\cal 
A}_0^{(23)} +m_2/m_1$ turns the final term into a mass term for ${\cal 
A}_0^{(23)}$.  Thus whether or not $m_1$ is non-zero, the theory actually 
has just two cosmological terms, and admits 2-charge domain wall solutions 
using $e^{\vec a_{1234}\cdot\vec\phi}$ and either 
$e^{\vec b_{124}\cdot \vec\phi}$ or $e^{\vec b_{134}\cdot \vec\phi}$.  It is 
easily verified from the equations in the appendix that in both cases
(\ref{dot}) is satisfied, and therefore there exist domain-wall solutions 
carrying two charges, namely $m_1$ and $m_3$, or $m_2$ and $m_3$.  
Configurations in the seven-dimensional theory are reinterpreted from the 
eleven-dimensional point of view as $U(1)$ bundles over $T^3$.  This can be 
easily seen from the structure of the vielbeins in the four internal 
directions, which take the form
\bea
&&e^1\sim (dz_1 + m_1\, z_4\, dz_2 + m_2\, z_4\, dz_3 +
{\cal A}_1^{(1)} + {\cal A}_0^{(1j)}\, dz_j) \, \qquad j > 1 \ ,\nn\\
&&e^i\sim (dz_i +{\cal A}_1^{(i)} + {\cal A}_0^{(ij)}\, dz_j) \, 
\qquad j > i>1 \ ,
\eea
where we have omitted the exponential dilatonic prefactors.
In the domain wall solutions themselves, only one charge appears in the 
vielbeins (the other is carried by the 4-form field strength).  This is 
manifest if $m_1=0$; if on the other hand $m_1\ne 0$ we will have ${\cal 
A}_0^{(23)}= m_2/m_1$, and the charge is associated with the single 
potential $z_4 (m_1\, dz_2 + m_2\, dz_3)$.

     More complicated examples of massive supergravities can be obtained 
where the fibres of the internal space are higher-dimensional tori $T^n$
with $n > 1$.  For example, there is a seven-dimensional massive theory
obtained by subjecting ${\cal A}_0^{(13)}$ and ${\cal A}_0^{(23)}$ to \ss
reductions from $D=8$.  From the eleven-dimensional point of view, 
configurations in $D=7$ are then interpreted in terms of $T^2$ bundles over 
$T^2$, as can be seen from the form of the internal vielbeins
\bea
&&e^1\sim (dz_1 + m_1\, z_4\, dz_3 +\cdots )\ , \qquad 
e^2\sim (dz_2 + m_2\, z_4\, dz_3 +\cdots)\ ,\nn\\
&&e^3\sim (dz_3 +\cdots) \ ,\qquad e^4\sim (dz_4 +\cdots)\ ,
\eea
where the ellipses represent the regular terms constructed from the 
seven-dimensional ${\cal A}_1^{(i)}$ and ${\cal A}_0^{(ij)}$ potentials, and 
we have again omitted the exponential dilatonic prefactors.
In this particular example, the seven-dimensional theory does not admit a 
2-charge domain wall solution, since the dot products of $\vec b_{134}$ and 
$\vec b_{234}$ do not satisfy (\ref{dot}).

    In the previous examples, multiple axions were subjected to \ss 
reductions, but all in the same step of dimensional reduction.  Other kinds 
of generalisation are possible in which the \ss ansatz is used at two or 
more different stages of a multi-step dimensional reduction procedure.
For example, we may set ${\cal A}_0^{(12)} =m_1\, z_3$ and ${\cal 
A}_0^{(14)} = m_2\, z_5$ to obtain a massive theory in $D=6$.  
Configurations in $D=6$ are described in terms of $T^1$ bundles over $T^4$ 
in the internal space.  In fact this example admits a 2-charge domain wall 
solution, since $\vec b_{123}$ and $\vec b_{145}$ do satsify the condition 
(\ref{dot}).  As a final example, we may construct a massive theory in $D=4$ 
by setting 
\be
{\cal A}_0^{(12)}= m_1\, z_5 \ ,\qquad 
{\cal A}_0^{(16)}= m_2\, z_7 \ ,\qquad
{\cal A}_0^{(35)}= m_3\, z_6 \ ,\qquad
{\cal A}_0^{(45)}= m_4\, z_7 \ ,
\ee
whose internal space when lifted back to eleven dimensions has the structure 
of a $T^3$ bundle over $T^4$.  This example allows a 4-charge domain wall 
solution.

\subsection{Properties of the internal space}

In this section we should like to clarify some interesting properties of 
the internal spaces arising in different compactification schemes. It
turns out that
there are a number of extra conditions to be imposed on the structure
of the internal space in order to have a consistent theory.  These additional
requirements involve the periodicities of the internal coordinates,
and the topological structure of the compactifying space. 

As our first example let us consider the oxidation to $D=11$ of the
domain wall solution obtained in section 2.1:
\be
ds_{11}^2=dx^\mu\, dx_\mu +Hdy^2+H(dz_2^2+dz_3^2)+H^{-1}(dz_1+m\, 
z_3\, dz_2)^2\ .
\ee
As we pointed out, it has the structure of a $U(1)$ principal fibre bundle, 
where $z_1$ is the fibre coordinate, while $z_2$ and $z_3$
are the coordinates on the base manifold, which in this particular case is
a 2-torus. One can define a connection form 
\be
{\cal A}_1^{(1)}=dz_1+m\, z_3\, dz_2 \label{con}.
\ee
The important property of this form is that it is globally
defined. Thus not 
only is it invariant under coordinate transformations, but also it
must give consistent
periodicities for the internal coordinates.  The internal space is
left invariant by
the following constant coordinate shifts: 
$z_i\rightarrow z_i+L_i$, where $L_i$ are the periods of 
corresponding compactifying  coordinates $z_i$. For instance, if one 
shifts $z_3$ in formula
(\ref{con}) by $L_3$, then in order to keep the whole expression invariant
this transformation has to be accompanied by a shift of $z_1$ that is given by
\be
z_1\rightarrow z_1+m\,  L_3 \, z_2\ .\label{shift} 
\ee
It is clear that above expression makes sense only if the right and left-hand
sides have the same transformation properties under the shift symmetry. 
In turn, this means that $L_1$ is non-trivially related to $L_2$ and $L_3$:
\be
L_1=m\,  L_2 \,  L_3\ .\label{R-rel1}
\ee  

   Similar relations can be established for other compactification
schemes.  In fact in general, the period $L$ for a coordinate $z$
whose compactification yields the Kaluza-Klein 2-form field strength
${\cal F}_2$ will be
\be
L = \int {\cal F}_2\ .
\ee
Let us consider some of the other exampls that were introduced
in section 2.  We already discussed the massive theory involving the 
simultaneous Scherk-Schwarz
reduction of the three axions ${\cal A}_0^{(12)}$, ${\cal A}_0^{(13)}$
and ${\cal A}_0^{(123)}$. It was shown that the theory admits a
2-charge domain wall solution, which in turn can be reinterpreted from
the $D=11$ point of view as a $U(1)$ principal bundle over $T^3$. It
is manifest from the structure of the vielbeins in the compact directions that
\bea
&&e^1\sim (dz_1+m_1\, z_4\, dz_2+m_2\, z_4\, dz_3+\dots)\ ,\nn\\
&&e^i\sim(dz_i+\dots), \qquad i>1\ ,
\eea
where we have omitted regular terms having no relevance to our present
discussion.  Analogous periodicity and consistency arguments can be given 
in this case also, leading to the following relations
\be
L_1=m_1\,  L_2 \, L_4 =  m_2\, L_3\,  L_4 \ .
\ee

   For a final example,  we shall consider an internal space with a
more complicated structure.
This comes from the seven-dimensional theory obtained by 
performing a \ss reduction of  ${\cal A}_0^{(13)}$ and 
${\cal A}_0^{(23)}$ in $D=8$. From the
eleven-dimensional point of view, all configurations in $D=7$ have the same
structure for their internal spaces, namely $T^2$ bundles over $T^2$,
as can be clearly seen from the form of vielbeins
\bea
&&e^1\sim(dz_1+m_1\, z_4\, dz_3+\dots)\ , \qquad e^2\sim(dz_2+m_2\,
z_4\, dz_3+\dots),\nn\\
&&e^3\sim(dz_3+\dots)\ , \qquad e^4\sim(dz_4+\dots)\ ,
\eea
where as always we present only the relevant terms and coefficients.
All the previous considerations go through without change here too.  Thus 
similar arguments imply the following relation between the periods of
the compactification coordinates
\be
L_1= m_1\, L_3 \, L_4, \qquad L_2= m_2\, L_3 \, L_4\ . \label{R-rel}
\ee

  To avoid confusion, it is should be emphasised that the 
compactification periods we are discussing here are not the same as
the physical sizes of the compact internal dimensions.  These are
related to the periods by metric factors involving the dilatonic
scalars, and so the actual size of the internal space is determined by the 
solution, rather
then fixed at will or by some compactification scheme.  But relations
of the form  (\ref{R-rel}) do have physical consequences, in that they impose
quantisation conditions on the mass parameters. 
To clarify this point, let us go 
back to the expresssion (\ref{R-rel1}). This formula fixes value of
mass parameter in terms of periods of the compactifying coordinates, 
\be
m=\fft{L_1}{L_2\, L_3}\ .
\ee
The above relation essentially  means that the mass parameter is determined,
once the compactification periods are chosen.  However, a more careful
analysis shows that this is not precisely correct.
Indeed, the assumption that the mass is simply proportional to $m_0$, where
$m_0=L_1/(L_2 \, L_3)$, with some integral coefficient, is good enough. It does
not break any invariance, and it is consistent with the  restrictions we have
imposed. Obviously this argument is valid in all examples we have given in the
paper. For instance from (\ref{R-rel}) it follows that
\be
m_1= n_1\, \fft{L_1}{L_2\, L_4}\ , \qquad m_2=n_2\, \fft{L_1}{L_3\,
L_4}\ , 
\ee
where $n_1$ and $n_2$ are integers. 

         Thus we saw that the the domain wall charges are discretised.
This discretisation of the charges are purely due to the classical
effect; it is quite different from the discretisation of usual
$p$-brane charges, where the existence of electric and magnetic pair
will suffer a Dirac quantisation condition.  Domain walls do not have
electric duals (they would be $(-2)$-branes).

\section{Fibre bundles and generalised reductions}

     In the previous section, we discussed classes of generalised
reductions in which axions with global abelian shift symmetries were
given a linear dependence on the compactification coordinate.  In
fact, as we shall explain later, one can make use of any global
symmetry to allow the dependence of certain fields on the coordinates
of the compactifying space.  The simplest example that goes beyond the
cases that we discussed previously is when the global symmetry is
associated with a shift symmetry of a dilatonic scalar field.  We
shall present an example of such a generalised reduction in the next
section. Before we go into detailed discussion of this example it is
worthwhile to describe some general ideas behind all generalised
reductions based on global symmetries.

   Let us suppose that we have a theory in $D+1$-dimensional space-time
with a Lagrangian that is left invariant by some global group $G$ of
transformations.   For simplicity, let us consider as an example a
theory comprising a set of scalar
fields $\phi^i$ in some representation of the group $G$, where the
the Lagrangian is left invariant by the transformation
\be
\hat\phi\rightarrow g\, \hat\phi\ .
\ee
Here $g$ is an element of the global symmetry group $G$, and
$\hat\phi$ is a vector with components
$\phi^i$. One can translate this picture into the language of fibre
bundles as follows.
The group $G$ becomes the structure group acting on sections 
(\ie the $\phi^i$ fields) of a
trivial fibre bundle with ($D+1$)-dimensional space-time as a base manifold.
The bundle is trivial because Minkowski space-time can be covered by
one coordinate chart.
The possibility of having a non-trivial bundle arises when one
compactifies one of the  spatial
coordinates on a circle. Now, the space-time $M_{D+1}$ decomposes into the
product  $S^1\times M_D$, which certainly can be used to give a 
non-trivial twist in the fibres.   Thus we may consider the situation where
the fields $\phi^i$ are not globally defined over $S^1$, but rather
are related by some element of the group $G$ when one goes from one 
coordinate chart to another. 

     The way to implement this is to introduce the $z$-dependent
factor $h(z)$ for the fields $\hat\phi$ in $D+1$ dimensions, so that
instead of making the trivial Kaluza-Klein ansatz
$\hat\phi(x,z)=\hat\phi(x)$, where $z$ are the compactifying
coordinates, we impose $\hat\phi(x,z)=h(z)\,
\hat\phi(x)$, where $h(z)$ satisfies $h(z_{(2)})=g\, h(z_{(1)})$. Here
$g$ is some group element in $G$, and $z_{(1)}$ and $z_{(2)}$ are
respectively the coordinates on the first and second coordinate charts
on $S^1$. In such a case one has two locally-defined sections
$\hat\phi_{(1)}$ and $\hat\phi_{(2)}$, related to one another on the
intersection of the two charts by the transformation
$\hat\phi_{(2)}=g\, \hat\phi_{(1)}$.  As we shall see, typically $g$
depends on parameters characterising the fibre bundle, as does the
resulting $D$-dimensional theory.  In more general examples, the
compact space need not be $S^1$. Later, we shall study
compactifications where it is $T^n$, or more complex spaces.  We shall
now present some examples to make this abstract discussion a little
more concrete.

\section{Generalised reductions for Cremmer-Julia symmetries}

\subsection{Scherk-Schwarz reduction of the dilaton}

Unlike the shift symmetries of axions, these transformations 
must be accompanied by appropriate rescalings of the higher-degree 
tensor fields 
in the theory.  Consider, for example, the type IIA theory in ten 
dimensions, for which the bosonic Lagrangian takes the form
\be
{\cal L} = eR -\ft12 e\, (\del\phi_1)^2 -\ft1{48} e \, e^{-\ft12\phi_1}\,
F_4^2 -\ft1{12} e\, e^{\phi_1}\, (F_3^{(1)})^2 -\ft14 e\, e^{-\ft32 \phi_1}\, 
({\cal F}_2^{(1)})^2 + \ft12 \ast(dA_3\wedge dA_3\wedge A_2^{(1)}) \ ,
\label{d10lag}
\ee
where $F_4=dA_3 -d A_2^{(1)}\wedge {\cal A}_1^{(1)}$, $F_3^{(1)} =dA_2^{(1)}$ 
and ${\cal F}_2^{(1)}=d{\cal A}_1^{(1)}$.  This is invariant under the
following global shift transformation of the dilaton: 
\be
\phi_1\longrightarrow \phi_1 + c\ ,
\qquad {\cal A}_1^{(1)}\longrightarrow e^{\ft34 
c}\, {\cal A}_1^{(1)}\ ,\qquad A_2^{(1)}\longrightarrow e^{-\ft12 c}\, 
A_2^{(1)}\ ,\qquad
A_3\longrightarrow e^{\ft14 c}\, A_3\ .\label{glob1}
\ee

     Clearly one can transform to variables in which the dilaton is covered
by a derivative everywhere, by redefining each gauge potential $A$, with
kinetic term $e^{a\phi}\, F^2$, according to $A\rightarrow e^{-\ft12
a\phi}\, A$.  Equivalently, we may simply make an appropriate generalised
reduction ansatz on the original variables, of the form 
\bea
\phi_1(x,z) &\longrightarrow& \phi_1(x) + m\, z\ ,\nn\\
{\cal A}_1^{(1)}(x,z) &\longrightarrow& 
e^{\ft34m\, z} \Big({\cal A}_1^{(1)}(x) + 
{\cal A}_0^{(12)}(x)\wedge dz\Big)\ ,\nn\\
A_2^{(1)}(x,z) &\longrightarrow& e^{-\ft12m\, z} \Big(A_2^{(1)}(x) + 
A_1^{(12)}(x)\wedge dz\Big)\ ,\label{ssred1}\\
A_3(x,z) &\longrightarrow& e^{\ft14m\, z} \Big(A_3(x) + 
A_2^{(2)}(x)\wedge dz\Big)\ .\nn
\eea
This corresponds to setting the global parameter $c$ in (\ref{glob1})
equal to $m\, z$, and then making the usual $z$-independent reduction
ansatz on the transformed fields.  Because of the shift symmetry, it
is guaranteed that if we substitute (\ref{ssred1}), together with the
standard Kaluza-Klein ansatz for the metric, into the ten-dimensional
equations of motion following from (\ref{d10lag}), the resulting
equations will be independent of the compactification coordinate $z$,
and will correspond to a consistent truncation to $D=9$.  Accordingly,
we may instead consistently substitute the ansatz into the
ten-dimensional Lagrangian, to obtain the nine-dimensional result.
This will be very similar to the form of the usual massless
supergravity, given by (\ref{dgenlag}) with $D=9$, except that now
some of the field strengths have acquired extra terms involving the
mass parameter $m$, and also there will be a cosmological term, given
by
\be
{\cal L}_{\rm cosmo} = -\ft12 m^2\, e^{\ft4{\sqrt7}\,\phi_2}
\ .\label{9cosmo}
\ee  
The field strengths have the form
\bea
{\cal F}_1^{(12)} &=& d{\cal A}_0^{(12)} -\ft34 m\, {\cal A}_1^{(1)} \ ,
\qquad {\cal F}_2^{(1)} = d{\cal A}_1^{(1)} +\cdots\ ,\nn\\
F_2^{(12)} &=& dA_1^{(12)} + \ft12 m\, A_2^{(1)} \ ,
\qquad F_3^{(1)} = dA_2^{(1)} +\cdots \ ,\nn\\
F_3^{(2)} &=& dA_2^{(2)} -\ft14 m\, A_3 + \cdots\ ,\qquad
F_4=dA_3 + \cdots\ ,\nn\\
{\cal F}_2^{(2)} &=& d{\cal A}_1^{(2)}\ ,
\eea
where the ellipses represent higher-order terms associated with the 
Kaluza-Klein modifications.  It is clear from these expressions that the 
fields ${\cal A}_1^{(1)}$, $A_2^{(1)}$ and $A_3$ become massive, eating in 
the process the fields ${\cal A}_0^{(12)}$, $A_1^{(12)}$ and $A_2^{(2)}$
respectively.  It is interesting to note also that the coupling of the 
dilatonic scalar $\phi_2$ in the cosmological term (\ref{9cosmo}) is 
characterised in terms of the parameter $\Delta$, defined in
\cite{stainless}, by $\Delta=0$.   

\subsection{$SL(2,\R)$ Scherk-Schwarz reduction in type IIB}

     In this subsection, we shall discuss a Scherk-Schwarz reduction
that makes 
use of the global $SL(2,R)$ symmetry of the type IIB supergravity theory.  
This illustrates a procedure that can be applied in a more general context, 
for any global symmetry.  

     Following the notation in \cite{sch},  we may write a Lagrangian for a 
bosonic subsector of the type IIB theory, where the self-dual 5-form is set 
to zero.  Since this is a singlet under $SL(2,R)$, it is not in any case of 
relevance in our discussion.  The Lagrangian takes the form
\be
{\cal L} = e R +\ft14 e\, {\rm tr}(\del {\cal M}\, \del {\cal M}^{-1}) 
-\ft1{12} e\, H_3^T\,  {\cal M}\, H_3\ ,\label{2blag}
\ee
where $H_3=\pmatrix{H_3^{(1)}\cr H_3^{(2)}}$, with $H_3^{(1)}$ and 
$H_3^{(2)}$ being the NS-NS and R-R 3-forms. ${\cal M}$ is the matrix
\be
{\cal M} = e^\phi\, \pmatrix{|\tau|^2 & \chi\cr \chi & 1}\ ,
\ee
where $\tau = \chi + i\, e^{-\phi}$, and $\chi$ and $\phi$ are the axion and 
dilaton.  The Lagrangian is invariant under the global $SL(2,R)$ transformation
\be
{\cal M}\longrightarrow \Lambda\, {\cal M}\, \Lambda^T\ ,
\qquad B_2\longrightarrow (\Lambda^T)^{-1}\, B_2\ ,
\ee
where $H_3^{(i)} = dB_2^{(i)}$.  Let us introduce a matrix $\Lambda$, which is 
effectively the ``square root'' of ${\cal M}$, such that ${\cal M} = 
\Lambda^T\, \Lambda$.  One can, for example, take
\be
\Lambda = \pmatrix{e^{-\ft12\phi} & 0\cr \chi\, e^{\ft12 \phi} & 
e^{\ft12\phi} }\ .\label{lam}
\ee

     Let us now perform a generalised dimensional reduction on a circle, 
where $z$ dependence is introduced in the following way:
\bea
{\cal M}(x,z) &=& \lambda(z)^T\, {\cal M}(x)\, \lambda(z)\ .\nn\\
B_2(x,z) &=& \lambda(z)^{-1}\, \Big( B_2(x) + B_1(x)\wedge dz \Big)\ ,
\label{bred}
\eea
together with the standard Kaluza-Klein ansatz for the metric.  The 
$SL(2,R)$ matrix $\lambda(z)$ is required to have the following property:
\be
(\del_z\, \lambda(z))\, \lambda(z)^{-1}  = C\ ,
\ee
where $C$ is a constant matrix of the form
\be
C = \pmatrix{\ft12 m_1 & 0\cr m_2 & -\ft12 m_1}\ ,
\ee  
with $m_1$ and $m_2$ being arbitrary constants.  Substituting 
into (\ref{2blag}), we see that the ten-dimensional Lagrangian reduces to
\bea
{\cal L}_9 &=& eR -\ft12 e\, (\del\varphi)^2 +\ft14e\, 
{\rm tr}\Big( \del {\cal 
M}\, \del{\cal M}^{-1}\Big)  -\ft14e\, e^{-\ft4{\sqrt7}\varphi}\, {\cal 
F}_2^2 -\ft1{12} e\, e^{-\ft1{\sqrt7}\varphi}\, 
H_3^T\, {\cal M}\, H_3 \nn\\
&&-\ft14 e\, e^{\ft3{\sqrt7}\varphi}\, H_2^T\,
{\cal M}\, H_2 -\ft12 e\, e^{\ft4{\sqrt7}\varphi}\, {\rm tr}\Big(C^2 +
C^T\, {\cal M}\, C\, {\cal M}^{-1}\Big)\ ,\label{d92b}
\eea
where $\varphi$ and ${\cal F}_2=d{\cal A}_1$ come from the 
dimensional reduction of the metric.  The nine-dimensional 3-form and 2-form 
field strengths are given by
\be
H_3=dB_2 -H_2\wedge {\cal A}_1\ ,\qquad H_2=dB_1 -C\, B_2\ ,
\ee
showing that under appropriate choices for $C$, the fields $B_2$ become 
massive, eating $B_1$ in the process.
Note that the ``cosmological terms'' described by the final expression in 
(\ref{d92b}) take the form
\be
{\cal L}_{\rm cosmo} = -\ft12 m_1^2 e\, e^{\ft4{\sqrt7}\varphi} -\ft12 e\, 
(m_2+m_1\, \chi)^2\, e^{2\phi +\ft4{\sqrt7}\varphi}
\ee
in terms of $\phi$ and $\chi$.  In fact when $m_1\ne 0$, the latter term is 
actually a mass term for $\chi$.

     The $z$ dependence of the ten-dimensional scalar fields $\phi(x,z)$ and 
$\chi(x,z)$ is dictated by the structure of the $z$-dependent matrix 
$\lambda(z)$, together with the ansatz for ${\cal M}(x,z)$ given in 
(\ref{bred}).  This can be re-expressed in the form $\Lambda(x,z) = \Lambda(x)
\, \lambda(z)$, where ${\cal M}(x)=\Lambda(x)^T\, \Lambda(x)$.  The matrix 
$\lambda(z)$ can be written as
\be
\lambda(z) = e^{z C} = \pmatrix{e^{\ft12 m_1\, z} & 0\cr
       \fft{2 m_2}{m_1}\, \sinh(\ft12 m_1\, z) & e^{-\ft12 m_1\, z} }\ .
\ee        
 From this, it follows that the ten-dimensional fields must have $z$ dependence
given by
\bea
\phi(x,z) &=& \phi(x) - m_1\, z\ ,\nn\\
\chi(x,z) &=& e^{m_1\, z}\, \chi(x) + \fft{m_2}{m_1}\, (e^{m_1\, z} -1)
\ . \label{zdep}
\eea

\subsection{$SL(2,R)$ Scherk-Schwarz reduction in type IIA}

      If the type IIA theory is reduced to nine dimensions, it also 
has an $SL(2,R)$ global symmetry.  This can be used for performing
a \ss reduction, in the same way as we did for the type IIB theory
above.  The advantage
of looking at this example is that the $SL(2,R)$ can be given a clear
geometric origin, by lifting the theory up to eleven dimensions.  (The
analogous F-theory origin of the $SL(2,R)$ symmetry in the type IIB theory
is not so well understood.)  In the discussion that follows, we shall focus
on the subsector of the theory that is relevant for describing domain-wall
solutions in the \ss reduced eight-dimensional theory.  This subsector is
obtained from the \ss reduction of ${\cal L}= eR +\ft14e \, {\rm tr} 
(\del{\cal M}\, \del{\cal M}^{-1})$, exactly as in the type IIB example above.
Thus the relevant part of the eight-dimensional Lagrangian will be
\bea
{\cal L}_8 &=& eR -\ft12 e\, (\del\varphi)^2 +\ft14 e\, {\rm tr} 
(\del{\cal M}\, \del{\cal M}^{-1}) \nn\\
&&-\ft12 m_1^2 e\, e^{\sqrt{\ft73}\varphi} -\ft12 e\, 
(m_2+m_1\, \chi)^2\, e^{2\phi +\sqrt{\ft73}\varphi}\ .\label{d8lag}
\eea
This admits two domain wall solutions, corresponding to either $m_1=0$, 
$m_2\ne0$, or $m_2=0$, $m_1\ne0$.  The former has a dilaton coupling
described by $\Delta=4$, and has a solution of the standard form 
\cite{lpss,lpdomain}
\bea
ds_8^2 &=& H^{1/6}\, dx^\mu\, dx_\mu +  H^{7/6}\, dy^2\ ,\nn\\
e^\phi &=& H^{\sqrt{3/19}}\ ,\qquad e^{\varphi} = H^{\ft12\sqrt{7/19}}\ .
\label{sol1}
\eea
where $H=1+ m_2\, |y|$. The latter gives a solution with $\Delta=0$.
After oxidation back to $D=11$, the solution (\ref{sol1}) gives
rise to the metric 
\be
ds_{11}^2=ds_8^2+Hdy^2+Hdz_3^2+He^{m_1z_3}dz_2^2+H^{-1}e^{-m_1z_3}dz_1^2.
\ee
The three-dimensional manifold described by coordinates $(z_1, z_2, z_3)$ has 
the structure of an homogeneous space, as can be easily shown from the
fact that at a fixed value of the coordinate $y$ the vielbeins 
$e^1=\lambda^{-1} e^{-\ft{m_1}2 z_3} dz_1$, 
$e^2=\lambda e^{\ft{m_1}2 z_3}dz_2$, and $e^3=\lambda dz_3$ satisfy
\be
de^1=-\ft{m_1}{2\lambda}e^3\wedge e^1, \qquad de^2=\ft{m_1}{2\lambda} e^3\wedge
e^2, \qquad de^3=0.
\ee
The spin connection and curvature 2-form are therefore given by
\bea
&&\omega_{12} = 0\ ,\qquad \omega_{23} = \fft{m_1}{2\lambda}\, e^2\ ,
\qquad \omega_{31} = \fft{m_1}{2\lambda}\, e^2\ ,\nn\\
&&\Theta_{12} = \fft{m_1^2}{4\lambda^2}\, e^1\wedge e^2\ ,\qquad
\Theta_{23} = -\fft{m_1^2}{4\lambda^2}\, e^2\wedge e^3\ ,\qquad
\Theta_{31} = -\fft{m_1^2}{4\lambda^2}\, e^3\wedge e^1\ .\qquad
\eea
In this case, the metric is of type V in the Bianchi classification scheme.
It is worth mentioning that all solutions of eight-dimensional supergravity
theory will preserve the structure of this metric on the internal space.

\section{Generalised reduction from $D=11$}

     As another example of a more general kind of \ss reduction, we may
take eleven-dimensional supergravity as a starting point.  The bosonic
Lagrangian takes the form
\be
{\cal L} = \hat e\, \hat R -\ft1{48}\, e\, \hat F_4^2 - 
\ft16 *(\hat F_4\wedge \hat F_4\wedge \hat A_3)\ , \label{d11}
\ee
where $\hat F_4=d\hat A_3$.  The theory has an homogeneous global scaling
symmetry, under which the fields $\hat g_{\sst{MN}}$ and $\hat
A_{\sst{MNP}}$ undergo the constant rescalings
\be
\hat g_{\sst{MN}}\longrightarrow \lambda^2\, \hat g_{\sst{MN}}\ , \qquad
\hat A_{\sst{MNP}}\longrightarrow \lambda^3\, \hat A_{\sst{MNP}}\ .
\label{trombone}
\ee
Although the action is not invariant under this transformation, the
equations of motion, namely 
\bea
\hat R{\sst{MN}} &=& \ft1{12}\Big(\hat F_{\sst{MPQR}}\, \hat F_{\sst
N}{}^{\sst{PQR}} -\ft1{12} \hat F_4^2\, \hat g_{\sst{MN}}\Big)\ ,\nn\\
\hat\nabla_{\sst M}\, \hat F^{\sst{MNPQ}} &=& \ft1{1152}\,
\epsilon^{\sst{NPQ R_1\cdots R_8}}\, \hat F_{\sst{R_1\cdots R_4}}\, 
\hat F_{\sst{R_5\cdots R_8}}\ ,\label{d11eom}
\eea
are invariant, by virtue of the fact that (\ref{trombone}) gives
an homogeneous scaling of the entire action.  Note that the equation of
motion for $\hat F_4$ may be written more elegantly as
\be
d*\hat F_4 =\ft12 \hat F_4\wedge \hat F_4\ .\label{elf}
\ee

     Let us now perform a
dimensional reduction to $D=10$ in which the eleven-dimensional fields are
allowed dependences on the compactifying coordinate $z$ of the form
\bea
d\hat s^2 &=& e^{2mz+\ft16\varphi}\, ds^2 + e^{2mz -\ft43\varphi}
(dz+{\cal A}_1)^2\ ,\nn\\
\hat A_3 &=& e^{3mz}\, A_3 + e^{3mz}\, A_2\wedge dz\ ,\label{d11ss}
\eea
where the ten-dimensional fields on the right-hand sides are independent of
$z$.  It is evident that if these expressions are substituted into the
eleven-dimensional equations of motion the $z$-dependence will cancel,
owing to the global scaling invariance (\ref{trombone}),  and thus we will
obtain a set of ten-dimensional equations of motion that are a consistent
truncation of the eleven-dimensional equations.  

     The computations are facilitated by first performing a Weyl rescaling
in $D=11$, and relating the Ricci curvature $\hat R_{\sst{MN}}$ of the
metric $\hat g_{\sst{MN}}$ to that of a Weyl rescaled metric $\tilde
g_{\sst{MN}}$ for which $\hat g_{\sst{MN}} = e^{2\sigma}\, 
\tilde g_{\sst{MN}}$.  In $n$ dimensions, a simple calculation gives
\be
\hat R_{\sst{MN}} =\widetilde R_{\sst{MN}} +
(n-2)(\del_{\sst M}\sigma\, \del_{\sst N}\sigma -\, \widetilde\nabla_{\sst
M}\del_{\sst N}\sigma -\tilde g^{\sst{PQ}}\, \del_{\sst P}\sigma\, \del_{\sst
Q}\sigma\, \tilde g_{\sst{MN}} ) -\square\sigma\, \tilde g_{\sst{MN}}
\ .\label{weyl}
\ee
With the help of this equation, it is straightforward to substitute the
reduction ans\"atze (\ref{d11ss}) into the eleven-dimensional equations of
motion (\ref{d11eom}), to obtain the following ten-dimensional equations of
motion:
\bea
&&\square\varphi = -\ft38 e^{-\fft32\varphi}\, {\cal F}_2^2 +\ft1{12}
e^\varphi\, F_3^2 -\ft1{96} e^{-\fft12\varphi}\, F_4^2 +\ft{27}2 m^2\,
{\cal A}_\mu\, {\cal A}^\mu + 9m\, {\cal A}^\mu\del_\mu\varphi -\ft32 m\,
\nabla_\mu{\cal A}^\mu\ ,\nn\\
&&R_{\mu\nu} -\ft12 R\, g_{\mu\nu}= \ft12(\del_\mu\varphi\,
\del_\nu\varphi -\ft12 (\del\varphi)^2\, g_{\mu\nu}) 
+\ft12 e^{-\fft32\varphi}\, ({\cal F}_{\mu\rho}\, {\cal F}_\nu{}^\rho 
-\ft14{\cal F}_2^2\, g_{\mu\nu})\nn\\
&&\qquad\qquad\qquad\qquad
 +\ft1{4} e^\varphi\, (F_{\mu\rho\sigma}\,
F_\nu{}^{\rho\sigma} -\ft16 F_3^2\, g_{\mu\nu}) + 
\ft1{12} e^{-\fft12\varphi}\, (F_{\mu\rho\sigma\lambda}\,
F_\nu{}^{\rho\sigma\lambda} -\ft18 F_4^2\, g_{\mu\nu})\nn\\
&&\qquad\qquad\qquad\qquad
-9m^2\, ({\cal A}_\mu\, {\cal A}_\nu + 4{\cal A}_\rho\, {\cal
A}^\rho\, g_{\mu\nu}) - 36m^2\, e^{\fft32\varphi}\, g_{\mu\nu} \nn\\
&&\qquad\qquad\qquad\qquad
-\ft92 m\, (\nabla_\mu{\cal A}_\nu +\nabla_\nu{\cal A}_\mu
-2\nabla_\rho{\cal A}^\rho\, g_{\mu\nu})\nn\\
&&\qquad\qquad\qquad\qquad
 +\ft34m\, ({\cal
A}_\mu\del_\nu\varphi + {\cal A}_\nu\del_\mu\varphi -{\cal
A}^\rho\del_\rho \varphi\, g_{\mu\nu})\ ,\\
&&\nabla_\nu(e^{-\fft32\varphi}\, {\cal F}_\mu{}^\nu) = 12m\del_\mu\varphi
+18m^2\, {\cal A}_\mu + 9m\, e^{-\fft32\varphi}\, {\cal A}^\nu\, {\cal
F}_{\mu\nu} -\ft16 e^{-\fft12\varphi}\, F_{\mu\nu\rho\sigma}\,
F^{\nu\rho\sigma}\ ,\nn\\
&&\nabla^{\sigma}(e^{-\fft12\varphi}F_{\mu\nu\rho\sigma})= -6m\,e^{\varphi}\,
F_{\mu\nu\rho}+6m\, e^{-\fft12\varphi}\, 
{\cal A}^{\sigma}F_{\mu\nu\rho\sigma}
-\fft1{144}\epsilon_{\mu\nu\rho\sigma_1\dots\sigma_7}\,
F^{\sigma_1\sigma_2\sigma_3\sigma_4}\,F^{\sigma_5\sigma_6\sigma_7},\nn\\
&&\nabla^{\sigma}(e^{\varphi}F_{\mu\nu\sigma})= 6m\,e^{\varphi}\, 
{\cal A}^\sigma\, 
F_{\mu\nu\sigma}+\ft12\, e^{-\fft12\varphi}\, F_{\m\nu\sigma\rho}\, 
{\cal F}^{\sigma\rho} + \fft1{1152}
\epsilon_{\mu\nu\rho_1\dots\rho_8}\, F^{\rho_1\rho_2\rho_3\rho_4}
F^{\rho_5\rho_6\rho_7\rho_8}.\nn
\eea
Note that the last three equations can be concisely 
written using the language of differential forms as follows
\bea
&&d(e^{-\fft32\varphi}\, *{\cal F}_2)=-12m\,*d\varphi-18m^2\,*{\cal A}_1+
9m\,e^{-\fft32\varphi}\, {\cal A}_1\wedge *{\cal F}_2 - e^{-\fft12\varphi}\,
F_3\wedge {*F_4},\nn\\
&&d(e^{-\fft12\varphi}\, *F_4) = 6m \, e^\varphi\, 
*F_3 + 6m e^{-\fft12\varphi}\,
{\cal A}_1\wedge {*F_4} + F_4\wedge F_3\ ,\nn\\
&&d(e^\varphi\, {*F_3}) = 6m\, e^\varphi\, 
{\cal A}_1\wedge {* F_3} - e^{-\ft12\varphi}\, 
{\cal F}_2 \wedge {*F_4} + \ft12 F_4\wedge F_4\label{massth}. 
\eea
Here, the field strengths in $D=10$ are defined by
\bea
{\cal F}_2 &=& d{\cal A}_1\ ,\nn\\
F_3 &=& dA_2 -3m\, A_3\ ,\\
F_4 &=& dA_3 -dA_2\wedge{\cal A}_1 + 3m\, A_3\wedge {\cal A}_1\ .\nn
\eea
Note that the calculation of the field equations for $F_3$ and $F_4$ can be
simplified by first establishing the following {\it lemmata}:
\bea
\hat * X_n &=&(-1)^n\,  e^{(11-2n)mz -\fft16(n-1)\varphi} 
{*X_n}\wedge(dz+{\cal A}_1)\ ,\nn\\
\hat *(X_n\wedge (dz+{\cal A}_1)) &=& e^{(9-2n)mz + \fft16 (9-n)\varphi} \,
{*X_n}\ ,
\eea
where $X_n$ is any $n$-form living in the ten-dimensional base manifold,
and $\hat *$ and $*$ denote the Hodge duals in $D=11$ and $D=10$
respectively.  Our conventions for the definition of the Hodge dual
in $D$ dimensions are as follows:
\bea
*(dx^{\mu_1}\cdots dx^{\mu_p}) &=& \fft1{q!}\, 
\epsilon_{\nu_1\cdots \nu_q}{}^{\mu_1\cdots \mu_p}\, dx^{\nu_1}\cdots 
dx^{\nu_q}\ ,\nn\\
({*\omega})_{\mu_1\cdots\mu_q}&=&\fft1{p!}\, \epsilon_{\mu_1\cdots
\mu_q}{}^{\nu_1\cdots\nu_p}\, \omega_{\nu_1\cdots \nu_p}\ ,
\eea
where $q=D-p$, and $\omega$ is a $p$-form.  These imply that
\bea
{* *\omega} &=& (-1)^{pq-1}\, \omega\ ,\nn\\
({*\omega})\wedge\omega & =& -\fft1{p!}\, |\omega|^2\, d^{\sst D} x\ ,
\label{stars}
\eea
where $\omega$ is any $p$-form, $|\omega|^2 =
\omega_{\mu_1\cdots\mu_p}\, \omega^{\mu_1\cdots\mu_p}$, and $d^{\sst
D}x = (D!)^{-1}\, \epsilon_{\mu_1\cdots\mu_{\sst D}}\, dx^{\mu_1} 
\cdots dx^{\sst D}$ is the volume form.  (Had the signature been
Euclidean, rather than Lorentzian, the right-hand sides of the two
equations in (\ref{stars}) would each be multiplied by a further
$(-1)$ factor.)

    The massive ten-dimensional theory that we have obtained here is
essentially the same as the one obtained in \cite{hlw} (except that in
\cite{hlw} the 4-form field strength was set to zero before making the
reduction.)  The procedure by which we have obtained the
ten-dimensional theory is slightly different from the one used in
\cite{hlw}; in that paper, the eleven-dimensional supergravity
starting point was modified slightly, by solving the superspace
constraints with a {\it conformal} spin connection \cite{howe} rather than
the usual one.  The modified eleven-dimensional theory was then
reduced to ten dimensions using ordinary Kaluza-Klein reduction.  In
our case, by contrast, the eleven-dimensional starting point was the
usual one, and the dimensional reduction was modified to the
Scherk-Schwarz scheme.  As was observed in \cite{hlw}, the equations
of motion (\ref{massth}) for this massive theory cannot be derived
from a Lagrangian.  This  is understandable, in view of the fact that we 
derived the ten-dimensional equations by means of a truncation of the 
eleven-dimensional theory which, although consistent, could be performed only
at the level of the equations of motion, since it involved the use of a global 
transformation that is a symmetry of the equations of motion, but not of 
the action.
 
   It can be easily seen that the equations (\ref{massth}) possess the 
following set of local symmetries:
\bea
&&\varphi\rightarrow\varphi+\l\ ,\qquad 
{\cal A}_1\rightarrow{\cal A}_1-\fft2{3m}d\l\ ,\qquad
g_{\mu\nu}\rightarrow e^{-\fft32\l}\, g_{\mu,\nu}\ ,\nn\\ 
&&A_2\rightarrow e^{-2\l}A_2\ ,\qquad A_3\rightarrow 
e^{-2\l}(A_3-\fft2{3m}\,d\l\wedge A_2)\ ,
\eea
and also
\bea
&&A_2\rightarrow A_2+d\l_1+\l_2\ ,\nn\\
&&A_3\rightarrow A_3+\fft1{3m}d\l_2\ ,
\eea
where $\l$, $\l_1$ and $\l_2$ are arbitrary 0-form, 1-form and 2-form
parameters respectively.    Note in particular that the parameters
$\l$ and $\l_2$ describe St\"uckelberg shift symmetries of $\varphi$
and $A_2$ respectively.   This is a characteristic feature of massive
theories, where the extra 
longitudinal degrees of freedom of the massive fields (${\cal A}_1$
and $A_3$) are replaced by additional fields ($\varphi$ and $A_2$), 
together with St\"uckelberg gauge transformations that can be used to 
eliminate these extra fields.  Indeed, one can choose a gauge such that
$\varphi=0$ and $A_2=d\l_1$, by choosing $\l=-\varphi$ and $\l_2=-A_2$. In
this gauge, the set of equations (\ref{massth}) reduces to the form
\bea
&&3m\,d {*{\cal A}_1}=\ft32 {\cal F}_2\wedge*{\cal F}_2 +9 m^2
A_3\wedge * A_3 +\fft12 F_4\wedge * F_4 +27 m^2\,
{\cal A}_1\wedge * {\cal A}_1,\nn\\
&&d{*{\cal F}_2} = 
-18m^2\, {*{\cal A}_1} + 9m \,{\cal A}_1\wedge {*{\cal F}_2}+3m\,
A_3\wedge {* F_4} ,\nn\\
&&R_{\mu\nu} -\ft12 R\, g_{\mu\nu}= 
\ft12({\cal F}_{\mu\rho}\, {\cal F}_\nu{}^\rho 
-\ft14{\cal F}_2^2\, g_{\mu\nu})\nn\\
&&\qquad\qquad\qquad\qquad
 +\ft94 m^2 (A_{\mu\rho\sigma}\,
A_\nu{}^{\rho\sigma} -\ft16 A_3^2\, g_{\mu\nu}) + 
\ft1{12} (F_{\mu\rho\sigma\lambda}\,
F_\nu{}^{\rho\sigma\lambda} -\ft18 F_4^2\, g_{\mu\nu})\nn\\
&&\qquad\qquad\qquad\qquad
-9m^2\, ({\cal A}_\mu\, {\cal A}_\nu + 4{\cal A}_\rho\, {\cal
A}^\rho\, g_{\mu\nu}) - 36m^2\, g_{\mu\nu} \nn\\
&&\qquad\qquad\qquad\qquad
-\ft92 m\, (\nabla_\mu{\cal A}_\nu +\nabla_\nu{\cal A}_\mu
-2\nabla_\rho{\cal A}^\rho\, g_{\mu\nu})\nn\\
&&d {*F_4} = -18 m^2\,  
{*A_3} + 6m 
{\cal A}_1\wedge {*F_4}  -3m\, F_4\wedge A_3\ ,\nn\\
&&3m\,d {*A_3}= 18 m^2\, {\cal A}_1\wedge {*A_3} + 
{\cal F}_2\wedge {*F_4}  -\ft12 F_4\wedge F_4\label{massth1}, 
\eea
where now $F_4=dA_3 + 3m\, A_3\wedge {\cal A}_1$.
The first and last equations here, which were originally the equations
of motion for the fields $\varphi$ and $A_2$ that have been gauged to
zero, play the r\^ole of
constraints on the remaining fields, and are  not independent of the
other equations.  Specifically, the first equation can be obtained by 
differentiation of the second,  
accompanied by appropriate resubstitutions of the other field equations. 
Similarly, the final equation can be derived by differentiation
of the equation of motion for the 3-form potential, followed by
appropriate resubstitutions from the equations of motion. 
Eventually, therefore, we are left with a theory of massive 1-form and
3-form potentials, coupled to gravity.

     It is of interest to study solutions of this massive
ten-dimensional theory.  In \cite{hlw}, it was shown that, unlike the
Romans theory, this massive supergravity does do admit an 8-brane
domain-wall solution.  It was shown, however, that it admits a
time-dependent BPS solution that preserves $\ft12$ the supersymmetry
\cite{hlw}.  Here, we observe that there is another simple solution to
the equations of motion (\ref{massth1}), in
which the gauge fields $A_3$ and ${\cal A}_1$ are set to zero.  The only
remaining non-trivial equation in (\ref{massth1}) is the Einstein
equation, which reduces to
\be 
R_{\mu\nu} = 9m^2\, g_{\mu\nu}\ .\label{einst}
\ee
This admits a de Sitter spacetime solution.  We may re-interpret this
as a solution in eleven dimensions, by reversing the steps of the
generalised dimensional reduction.  Thus from (\ref{d11ss}), we find
that the metric in $D=11$ takes the form
\bea
ds_{11}^2 &=& e^{2mz}\, dz^2 + e^{2mz}\, ds_{10}^2\ ,\nn\\
       &=& d\rho^2 + m^2\, \rho^2\, ds_{10}^2\ ,
\eea
where $\rho=e^{mz}/m$.  Since the de Sitter metric $ds_{10}^2$ satisfies 
(\ref{einst}), it follows that it can be written as $ds_{10}^2=
m^{-2}\, d\Omega_{10}^2$, where $d\Omega_{10}^2$ is the metric on a
``unit radius'' de Sitter spacetime whose Riemann tensor is given by
$R_{\mu\nu\rho\sigma}=g_{\mu\rho}\, g_{\nu\sigma} -
g_{\mu\sigma}\, g_{\nu\rho}$, and so we have
\be
ds_{11}^2 = d\rho^2 + \rho^2\, d\Omega_{10}^2\ .\label{flat}
\ee
This is in fact a metric on flat space.  This can be seen from the
fact that if $d\Omega_{10}^2$ were the metric on the unit-radius
10-sphere, the metric (\ref{flat}) would be nothing but flat space in
hyperspherical polar coordinates.  The de Sitter metric is obtained
from the 10-sphere metric by Lorentzianising one of the coordinates,
but this will not affect the results for local curvature calculations,
and so the Riemann tensor for (\ref{flat}) vanishes.

\section*{Acknowledgement}

    We are grateful to E. Cremmer, B. Julia fand K.S. Stelle for discussions on
non-compact global symmetries in supergravity.

\appendix
\section{Lagrangian of $D$-dimensional maximal supergravity}

     In this appendix, we present the bosonic sector of the Lagrangian
for $D$-dimensional maximal supergravity, obtained by dimensional
reduction from $D=11$.  The notation and conventions are those of
\cite{lpsol}.
\bea
{\cal L} &=& eR -\ft12 e\, (\del\vec\phi)^2 -\ft1{48}e\, e^{\vec a\cdot
\vec\phi}\, F_4^2 -\ft{1}{12} e\sum_i
e^{\vec a_i\cdot \vec\phi}\, (F_3^{i})^2
-\ft14 e\, \sum_{i<j} e^{\vec a_{ij}\cdot \vec\phi}\, (F_2^{ij})^2
\nonumber\\
&& -\ft14e\, \sum_i e^{\vec b_i\cdot \vec\phi}\, ({\cal F}_2^i)^2
-\ft12 e\, \sum_{i<j<k} e^{\vec a_{ijk} \cdot\vec \phi}\,
(F_1^{ijk})^2 -\ft12e\, \sum_{i<j} e^{\vec b_{ij}\cdot \vec\phi}\,
({\cal F}_1^{ij})^2 + {\cal L}_{\sst{FFA}}\ ,\label{dgenlag}
\eea
where the ``dilaton vectors'' $\vec a$, $\vec a_i$, $\vec a_{ij}$,
$\vec a_{ijk}$,
$\vec b_i$, $\vec b_{ij}$ are constants that characterise the couplings of
the dilatonic scalars $\vec \phi$ to the various gauge fields.
They are given by \cite{lpsol}
\bea
&&F_{\sst{MNPQ}}\qquad\qquad\qquad\qquad\qquad\qquad\qquad\qquad
{\rm vielbein}\nonumber\\
{\rm 4-form:}&&\vec a = -\vec g\ ,\nonumber\\
{\rm 3-forms:}&&\vec a_i = \vec f_i -\vec g \ ,\nonumber\\
{\rm 2-forms:}&& \vec a_{ij} = \vec f_i + \vec f_j - \vec g\ ,
\qquad\qquad\qquad\qquad\qquad \,\,\, \,\vec b_i = -\vec f_i \ ,
\label{dilatonvec}\\
{\rm 1-forms:}&&\vec a_{ijk} = \vec f_i + \vec f_j + \vec f_k -\vec g
\ ,\qquad\qquad\qquad\qquad\vec b_{ij} = -\vec f_i + \vec f_j\ ,\nonumber \\
{\rm 0-forms:}&& \vec a_{ijk\ell} =\vec f_i +\vec f_j+\vec f_k +\vec f_\ell
-\vec g \ ,\qquad\qquad\quad\ \  \vec b_{ijk}=-\vec f_i +\vec f_j +\vec f_k\ ,
\nonumber
\eea
where the vectors $\vec g$ and $\vec f_i$ have $(11-D)$ components
in $D$ dimensions, and are given by
\bea
\vec g &=&3 (s_1, s_2, \ldots, s_{11-\sst D})\ ,\nonumber\\
\vec f_i &=& \Big(\underbrace{0,0,\ldots, 0}_{i-1}, (10-i) s_i, s_{i+1},
s_{i+2}, \ldots, s_{11-\sst D}\Big)\ ,\label{gfdef}
\eea
where $s_i = \sqrt{2/((10-i)(9-i))}$.  It is easy to see that they satisfy
\be
\vec g \cdot \vec g = \ft{2(11-D)}{D-2}, \qquad
\vec g \cdot \vec f_i = \ft{6}{D-2}\ ,\qquad
\vec f_i \cdot \vec f_j = 2\delta_{ij} + \ft2{D-2}\ .\label{gfdot}
\ee
We have also included the dilaton vectors $\vec a_{ijk\ell}$ and $\vec
b_{ijk}$ for ``0-form field strengths'' in (\ref{dilatonvec}), although they
do not appear in (\ref{dgenlag}), because they fit into the same general
pattern and they do arise in the generalised reduction procedures for
axions that we consider in section 2 in this paper 
\cite{ss1,bdgpt,clpst,lpdomain,llp,classp}.

The field strengths are given by
\bea
F_4 &=& \td F_4 - \gamma^{ij} \td F_3^i\wedge {\cal A}_1^j +\ft12
\gamma^{ik}\gamma^{j\ell} \td F_2^{ij} \wedge {\cal A}_1^k\wedge
{\cal A}_1^\ell - \ft16 \gamma^{i\ell}\gamma^{jm}\gamma^{kn}
\td F_1^{ijk}\wedge {\cal A}_1^\ell \wedge {\cal A}_1^m \wedge
{\cal A}_1^n\ ,\nonumber\\
F_3^i &=& \gamma^{ji}\td F_3^j + \gamma^{ji}\gamma^{k\ell} \td F_2^{jk}
\wedge {\cal A}_1^\ell + \ft12 \gamma^{ji}\gamma^{km}\gamma^{\ell n}
\td F_1^{jk\ell}\wedge {\cal A}_1^m \wedge {\cal A}_1^n\ ,\nonumber\\
F_2^{ij} &=& \gamma^{ki}\gamma^{\ell j} \td F_2^{k\ell} -
\gamma^{ki} \gamma^{\ell j} \gamma^{mn} \td F_1^{k\ell m}\wedge
{\cal A}_1^n\ ,\label{csterm}\\
F_1^{ijk} &=& \gamma^{\ell i} \gamma^{mj} \gamma^{nk} \td F_1^{\ell mn}
\ ,\nonumber\\
{\cal F}_2^i &=& \td {\cal F}_2^i - \gamma^{jk} \td {\cal F}_1^{ij} \wedge
{\cal A}_1^k\ ,\nonumber\\
{\cal F}_1^{ij} &=& \gamma^{kj} \td {\cal F}_1^{ik}\ ,\nn
\eea
where the tilded quantities represent the unmodified pure exterior
derivatives of the corresponding potentials, and $\gamma^{ij}$ is defined
by
\be
\gamma^{ij}=[(1+{\cal A}_0)^{-1}]^{ij}=\delta^{ij} -{\cal A}_0^{ij} +
{\cal A}_0^{ik}\, {\cal A}_0^{kj} +\cdots\ .\label{gam}
\ee
Recalling that ${\cal A}_0^{ij}$ is defined only for $j>i$ (and vanishes
if $j\le i$), we see that the series terminates after a finite number of
terms. The term ${\cal L}_{\sst{FFA}}$ in (\ref{dgenlag}) is the dimensional
reduction of the $\td F_4\wedge\td F_4\wedge A_3$ term in $D=11$, and is
given in lower dimensions in \cite{lpsol}.

\end{document}